\documentclass[12pt,a4paper]{article}
\usepackage{graphicx}
\newcommand{\vs}{\vspace{-0.25cm}}

\begin{document}
\begin{center}
{\Large
\textbf{Nuclear energy density functional from\\ chiral two- and three-nucleon 
interactions}\footnote{Work supported in part by BMBF, GSI and the DFG
cluster of excellence: Origin and Structure of the Universe.}} 

\bigskip
J.W. Holt, N. Kaiser and W. Weise\\

\bigskip

{\small Physik Department, Technische Universit\"{a}t M\"{u}nchen, 
D-85747 Garching, Germany\\

\smallskip

{\it email: nkaiser@ph.tum.de}}

\end{center}

\begin{abstract}
An improved density-matrix expansion is used to calculate the nuclear 
energy density functional from chiral two- and three-nucleon interactions. The 
two-body interaction comprises long-range one- and two-pion exchange contributions and a 
set of contact terms contributing up to fourth power in momenta. In addition we employ 
the leading order chiral three-nucleon interaction with its parameters $c_E, c_D$ and 
$c_{1,3,4}$ fixed in calculations of nuclear few-body systems. With this input the nuclear 
energy density functional is derived to first order in the two- and three-nucleon 
interaction. We find that the strength functions $F_\nabla(\rho)$ and $F_{so}(\rho)$ of 
the surface and spin-orbit terms compare in the relevant density range reasonably with 
results of phenomenological Skyrme forces. However, an improved description requires (at 
least) the treatment of the two-body interaction to second order. This 
observation is in line with the deficiencies in the nuclear matter equation of state 
$\bar E(\rho)$ that remain in the Hartree-Fock approximation with low-momentum two- and 
three-nucleon interactions.   
\end{abstract}

\bigskip


\section{Introduction}
The nuclear energy density functional approach is the many-body method of choice in 
order to calculate the properties of medium-mass and heavy nuclei in a systematic manner
\cite{reinhard,stone}. Parameterized non-relativistic Skyrme functionals \cite{sly,pearson}
as well as relativistic mean-field models \cite{serot,ring} have been widely and 
successfully used for such nuclear structure calculations. A complementary approach 
\cite{lesinski,drut,platter,drut2} focuses less on the fitting of experimental data, 
but attempts to constrain the analytical form of the functional and the values of its 
couplings from many-body perturbation theory and the underlying two- and three-nucleon 
interaction. Switching from conventional hard-core NN-potentials to low-momentum 
interactions \cite{vlowkreview,vlowk} is essential in this respect, because the nuclear 
many-body problem formulated in terms of the latter becomes significantly more 
perturbative. Indeed, second-order perturbative calculations including also three-body 
forces give already a good account of the bulk correlations in infinite nuclear matter 
\cite{achim,hebeler} and in doubly-magic nuclei \cite{roth}.

In many-body perturbation theory the contributions to the energy are written in terms of
density-matrices convoluted with the finite-range interaction kernels, and are therefore 
highly non-local in both space and time. In order to make such functionals numerically 
tractable in heavy open-shell nuclei it is desirable to develop simplified 
approximations for these functionals in terms of local densities and currents only. In 
such a construction the density-matrix expansion comes prominently into play as it 
removes the non-local character of the exchange (Fock) contribution to the
energy by mapping it onto a generalized Skyrme functional with 
density-dependent couplings. For some time the prototype for that has been 
the density-matrix expansion of Negele and Vautherin \cite{negele}, but recently
Gebremariam, Duguet and Bogner \cite{dmeimprov} have developed an improved 
version for spin-unsaturated nuclei. They have demonstrated that 
phase-space averaging techniques allow for a consistent expansion of both the 
spin-independent (scalar) part as well as the spin-dependent (vector) part of 
the density-matrix. The improved properties of the new phase-space averaged density-matrix
expansion have been  extensively studied via the Fock energy densities arising from 
schematic finite-range central, tensor and spin-orbit interactions for a large set of 
semi-magic nuclei (for further details see ref.\cite{dmeimprov}). 

In order to match with these new developments, the nuclear energy density functional as 
it emerges from chiral pion-nucleon dynamics has been recalculated in 
ref.\cite{efun}. This calculation has treated for isospin-symmetric (i.e. $N=Z$) 
nuclear systems  the effects from $1\pi$-exchange, iterated $1\pi$-exchange, and 
irreducible $2\pi$-exchange with intermediate $\Delta$-isobar excitations, including 
Pauli-blocking corrections up to three-loop order. Among other things, it has been found 
that the two- and three-body contributions to the spin-orbit coupling strength 
$F_{so}(\rho)$, as generated by $2\pi$-exchange, tend to cancel each other in the relevant 
density range $\rho \simeq 0.08\,$fm$^{-3}$, thus leaving room for the short-range 
nuclear spin-orbit interaction. The short-range components of the NN-interaction 
together with the constraints on them provided by the elastic scattering data (i.e. 
NN-phase shifts etc.) have not been considered explicitly in ref.\cite{efun}. 
Furthermore, a similar calculation of a microscopically constrained nuclear energy 
density functional derived from the chiral NN-potential at next-to-next-to-leading order 
(N$^2$LO) has been presented recently by Gebremariam, Bogner and Duguet in 
ref.\cite{microefun}. They have proposed that the density-dependent couplings 
associated with the pion-exchange interactions should be added to a standard Skyrme 
functional (with several adjustable parameters). In the sequel it has been demonstrated 
in ref.\cite{stoitsov} that this new energy density functional gives numerically stable 
results and that it exhibits a small but systematic reduction of the $\chi^2$-measure 
compared to standard Skyrme functionals (without any pion-exchange terms).
   
The purpose of the present paper is to derive a nuclear 
energy density functional with improved (chiral) two- and three-nucleon interactions.
We use for the two-body interaction the N$^3$LO chiral NN-potential which reaches 
at this order the quality of a high-precision NN-potential (in reproducing empirical 
NN-shifts and deuteron properties). The  N$^3$LO chiral potential consists of long-range 
one- and two-pion exchange terms and two dozen low-energy constants which parameterize 
the short-distance part of the NN-interaction. The latter contact potential written in 
momentum space gives the most general contribution up to fourth power in momenta. In the 
actual calculation we will use the version N$^3$LOW of the chiral NN-potential 
developed in refs.\cite{mach,n3low} by lowering the cut-off scale to $\Lambda= 414\,$MeV. 
This value coincides with the resolution scale inherent to the universal low-momentum 
NN-potential $V_{\rm low-k}$  \cite{vlowkreview,vlowk} to which all realistic NN-potentials 
flow after integrating out effects from momenta above the cut-off scale $\Lambda=2.1\,
$fm$^{-1}$. The low-momentum two-body interaction N$^3$LOW is supplemented by the leading 
order chiral three-nucleon interaction with its parameters $c_E$, $c_D$ and $c_{1,3,4}$ 
determined in calculations of nuclear few-body systems \cite{achim,3bodycalc}. Our paper 
is organized as follows. In section 2 we recall the basic features of the (improved) 
density-matrix expansion and the nuclear energy density functional for isospin-symmetric 
systems. In section 3 we present the two-body contributions to the various 
density-dependent strength functions $\bar E(\rho)$, $F_\tau(\rho)$, $F_d(\rho)$, 
$F_{so}(\rho)$ and $F_J(\rho)$, separately for the finite-range pion-exchange and the 
zero-range contact interactions. In section 4, we collect the corresponding analytical 
expressions for the three-body contributions grouped into contact ($c_E$), 
$1\pi$-exchange ($c_D$) and $2\pi$-exchange ($c_{1,3,4}$) terms. Section 5 is devoted to 
the discussion of our numerical results and sections 6 ends with a summary and an outlook. 

\section{Density-matrix expansion and energy density functional}
The starting point for the construction of an explicit nuclear energy density functional 
is the density-matrix as given by a sum over the energy eigenfunctions $\Psi_\alpha(\vec r
\,)$ representing occupied orbitals of the (non-relativistic) many-fermion system. 
According to Gebremariam, Duguet and Bogner \cite{dmeimprov} it can be expanded in 
relative and center-of-mass coordinates, $\vec a$ and $\vec r$, as follows:   
\begin{eqnarray} \sum_{\alpha}\Psi_\alpha( \vec r -\vec a/2)\Psi_\alpha^
\dagger(\vec r +\vec a/2) &=& {3 \rho\over a k_f}\, j_1(a k_f)-{a \over 2k_f} 
\,j_1(a k_f) \bigg[ \tau - {3\over 5} \rho k_f^2 - {1\over 4} \vec \nabla^2 
\rho \bigg] \nonumber \\ && + {3i \over 2a k_f} \,j_1(a k_f)\, \vec \sigma
\cdot (\vec a \times \vec J\,) + \dots\,,  \end{eqnarray}
with the spherical Bessel function $j_1(x) = (\sin x - x \cos x)/x^2$. The quantities 
appearing on the right hand side of eq.(1) are: the (local) nucleon density $\rho(\vec r\,) 
=2k_f^3(\vec r\,)/3\pi^2 =  \sum_\alpha \Psi^\dagger_\alpha(\vec r\,) \Psi_\alpha(\vec r\,)$, 
the (local) kinetic energy density $\tau(\vec r\,) =  \sum_\alpha \vec \nabla 
\Psi^\dagger_\alpha (\vec r\,) \cdot \vec \nabla \Psi_\alpha(\vec r\,)$ and the (local) 
spin-orbit density $ \vec J(\vec r\,) = i \sum_\alpha \vec \Psi^\dagger_\alpha(\vec r\,) 
\vec \sigma \times \vec \nabla \Psi_\alpha(\vec r\,)$. As shown in section 2 of 
ref.\cite{efun} the Fourier transform of the expanded density-matrix eq.(1) with respect to 
both coordinates $\vec a$ and $\vec r$ defines in momentum space a ''medium insertion'':  
\begin{eqnarray} \Gamma(\vec p,\vec q\,)& =& \int d^3 r \, e^{-i \vec q \cdot
\vec r}\,\bigg\{ \theta(k_f-|\vec p\,|) +{\pi^2 \over 4k_f^4}\Big[k_f\,\delta'
(k_f-|\vec p\,|)-2 \delta(k_f-|\vec p\,|)\Big] \nonumber \\ && \times \bigg( 
\tau - {3\over 5} \rho k_f^2 - {1\over 4} \vec \nabla^2 \rho \bigg) -{3\pi^2 
\over 4k_f^4}\,\delta(k_f-|\vec p\,|) \, \vec \sigma \cdot (\vec p \times 
\vec J\,)  \bigg\}\,,  \end{eqnarray}
for inhomogeneous many-nucleon systems characterized by the time-reversal-even fields 
$\rho(\vec r\,)$,  $\tau(\vec r\,)$ and $ \vec J(\vec r\,)$. Note that the 
delta-function $\delta(k_f-|\vec p\,|)$ in eq.(2) gives weight to the momentum-dependent 
NN-interactions only in the vicinity of the local Fermi momentum, $|\vec p\,|=k_f(
\vec r\,)$. 

Up to second order in spatial gradients (i.e. deviations from homogeneity) the 
energy density functional relevant for $N=Z$ even-even nuclei reads:
\begin{eqnarray} {\cal E}[\rho,\tau,\vec J\,] &=& \rho\,\bar E(\rho)+\bigg[\tau-
{3\over 5} \rho k_f^2\bigg] \bigg[{1\over 2M}-{k_f^2 \over 4M^3}+F_\tau(\rho)
\bigg] \nonumber \\ && + (\vec \nabla \rho)^2\, F_\nabla(\rho)+  \vec \nabla 
\rho \cdot\vec J\, F_{so}(\rho)+ \vec J\,^2 \, F_J(\rho)\,.\end{eqnarray}
Here, $\bar E(\rho)$ is the energy per particle of isospin-symmetric nuclear matter 
evaluated at the local nucleon density  $\rho(\vec r\,)$. The strength function 
$F_\tau(\rho)$ introduces an effective (density-dependent) nucleon mass $M^*(\rho)$ and it 
is related to the single-particle potential $U(p,k_f)$ as follows:
\begin{equation} F_\tau(\rho) = {1 \over 2k_f} {\partial U(p,k_f) \over \partial p}
\Big|_{p=k_f} = -{k_f \over 3\pi^2} f_1(k_f)\,,\end{equation}
with $\rho = 2k_f^3/3\pi^2$. The second equality establishes the equivalent relation to the 
spin and isospin independent p-wave Landau parameter $f_1(k_f)$. The strength function  
$F_\nabla(\rho)$ of the $(\vec \nabla \rho)^2$ surface term has the decomposition 
\cite{efun}:     
\begin{equation} F_\nabla(\rho) = {1\over 4}\, {\partial F_\tau(\rho)
\over  \partial \rho} +F_d(\rho) \,,\end{equation}
where $F_d(\rho)$ comprises all those contributions for which the $(\vec \nabla \rho
)^2$-factor originated directly from the momentum dependence of the interactions in an 
expansion up to order $\vec q^{\,2}$. Note that only the (fixed) nuclear matter piece 
$\theta(k_f-|\vec p\,|)$ of the density-matrix expansion goes into the derivation 
of the strength function  $F_d(\rho)$. The second to last term $\vec \nabla \rho \cdot\vec 
J\, F_{so}(\rho)$ in eq.(3) is responsible for the spin-orbit interaction in nuclei.  
The associated function $F_{so}(\rho)$ measures therefore the strength of the nuclear 
spin-orbit coupling.  

\section{Two-body contributions}
In this section we work out the two-body contributions to the various density-dependent 
strength functions which build up the nuclear energy density functional ${\cal E}[
\rho,\tau,\vec J\,]$ written in eq.(3). Ideally, one would like to use for this task 
the universal low-momentum NN-potential $V_{\rm low-k}$ \cite{vlowk}. However, it 
is generally given in terms of (off-shell) partial wave matrix elements which makes 
its application to the density-matrix expansion rather cumbersome. An explicit 
representation of the momentum space NN-potential in terms of spin- and 
isospin-operators is much better suited for this purpose. For this reason we use (as a 
substitute for $V_{\rm low-k}$) the chiral NN-potential N$^3$LOW developed in 
refs.\cite{mach,n3low} by lowering the cut-off scale to $\Lambda= 414\,$MeV. This value of 
$\Lambda$ coincides with the resolution scale inherent to the universal low-momentum 
NN-potential $V_{\rm low-k}$. The finite-range part of the  N$^3$LOW chiral NN-potential 
consists of one- and two-pion exchange pieces which can be summarized in the 
form:\footnote{Note that we associate here the tensor interaction with the operator 
$\vec \sigma_1 \cdot \vec q \,\, \vec\sigma_2 \cdot \vec q$. This operator splits as 
$(q^2/3) [S_{12}(\hat q)+ \vec \sigma_1 \cdot  \vec\sigma_2]$ into the genuine tensor 
operator $S_{12}(\hat q)$ and a spin-spin piece.}
\begin{eqnarray} V_{NN}^{(\pi)} &=& V_C(q) + \vec \tau_1 \cdot \vec \tau_2\, W_C(q) +
\big[V_S(q) + \vec \tau_1 \cdot \vec \tau_2\, W_S(q)\big] \, \vec \sigma_1 \cdot 
\vec\sigma_2 \nonumber \\ && +\big[V_T(q) + \vec \tau_1 \cdot \vec\tau_2\,
W_T(q) \big]\,\vec \sigma_1 \cdot \vec q \,\, \vec\sigma_2 \cdot \vec q  
\nonumber \\ && +\big[V_{SO}(q) + \vec \tau_1 \cdot \vec\tau_2\,W_{SO}(q)\big]\, 
i (\vec \sigma_1+\vec\sigma_2)\cdot (\vec q \times \vec p\,) \,, \end{eqnarray}
where $\vec q$ denotes the momentum transfer and $\vec p$ the center-of-mass momentum. 
As usual, $\vec \sigma_{1,2}$ and $\vec \tau_{1,2}$ are the spin- and isospin operators of 
the two nucleons. A special and simplifying feature  of $V_{NN}^{(\pi)}$ is that all the 
occurring potentials $V_C(q),\dots, W_{SO}(q)$ depend only on the momentum transfer $q$ 
and that a quadratic spin-orbit component $\sim \vec \sigma_1\cdot (\vec q \times \vec 
p\,)\,\vec \sigma_2\cdot (\vec q \times \vec p\,)$ is absent. The relativistic 
$1/M^2$-correction to the $2\pi$-exchange \cite{relcor} which does (partially) not 
share this property is so small that it can be safely neglected. In order to specify our 
sign and normalization convention, we give also the explicit expression for the 
$1\pi$-exchange, $W_T^{(1\pi)}(q) =-(g_A/2f_\pi)^2(m_\pi^2+q^2)^{-1}$, with the parameters 
$g_A=1.3$, $f_\pi=92.4\,$MeV and $m_\pi=138\,$MeV.

\begin{figure}
\begin{center}
\includegraphics[scale=0.43,clip]{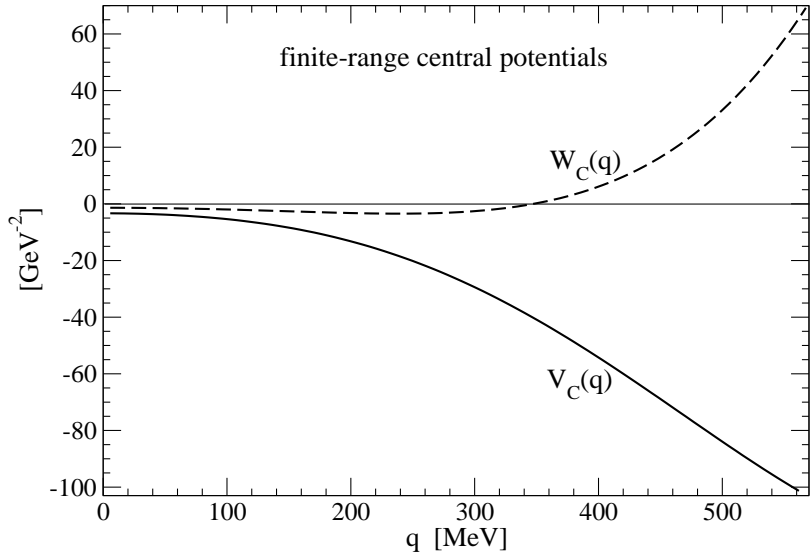}
\end{center}
\vspace{-.8cm}
\caption{Finite-range isoscalar and isovector central potentials extracted from 
N$^3$LOW \cite{n3low}.}
\end{figure} 

\begin{figure}
\begin{center}
\includegraphics[scale=0.43,clip]{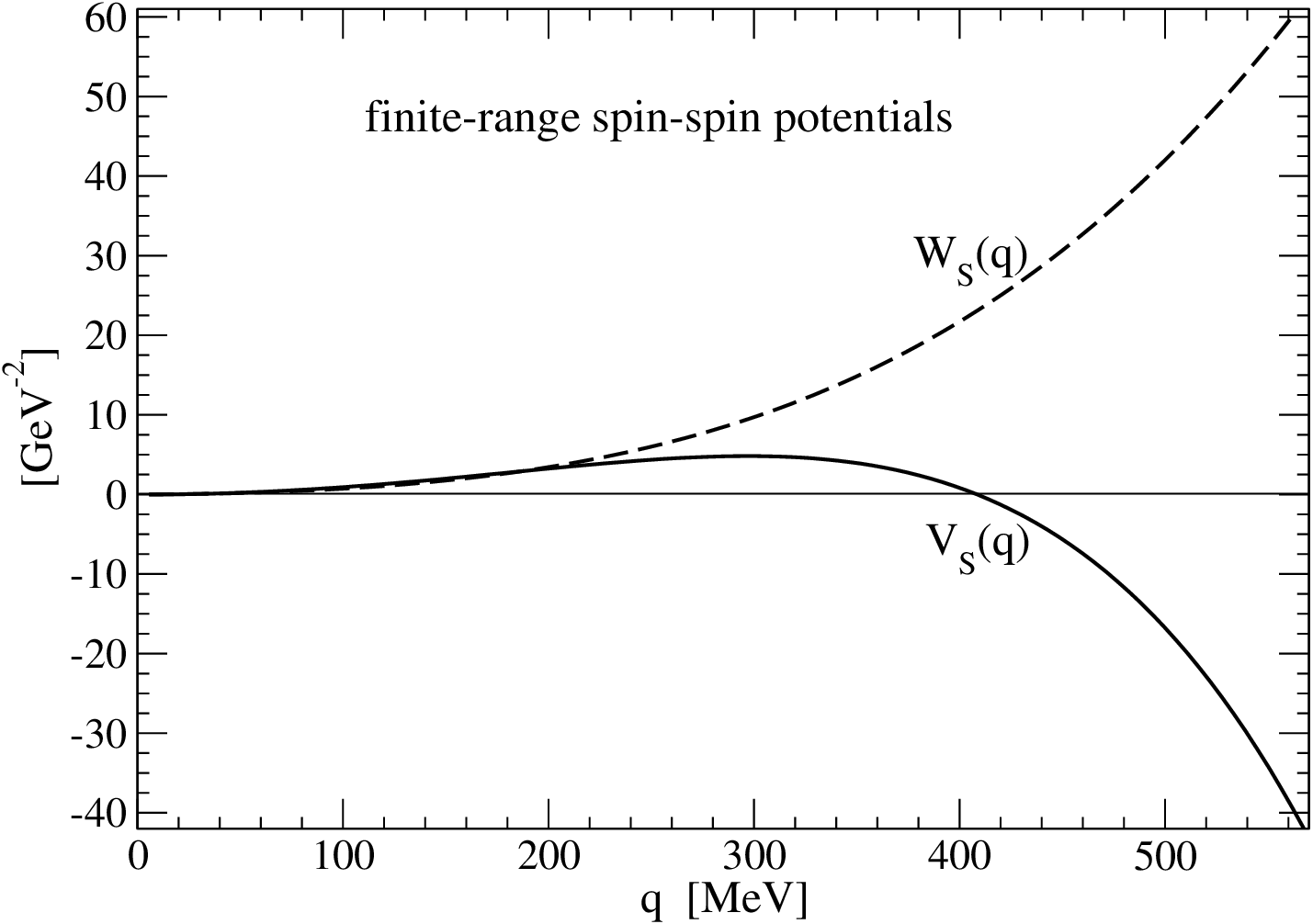}
\end{center}
\vspace{-.8cm}
\caption{Finite-range isoscalar and isovector spin-spin potentials extracted from 
N$^3$LOW \cite{n3low}.}
\end{figure}

\begin{figure}
\begin{center}
\includegraphics[scale=0.43,clip]{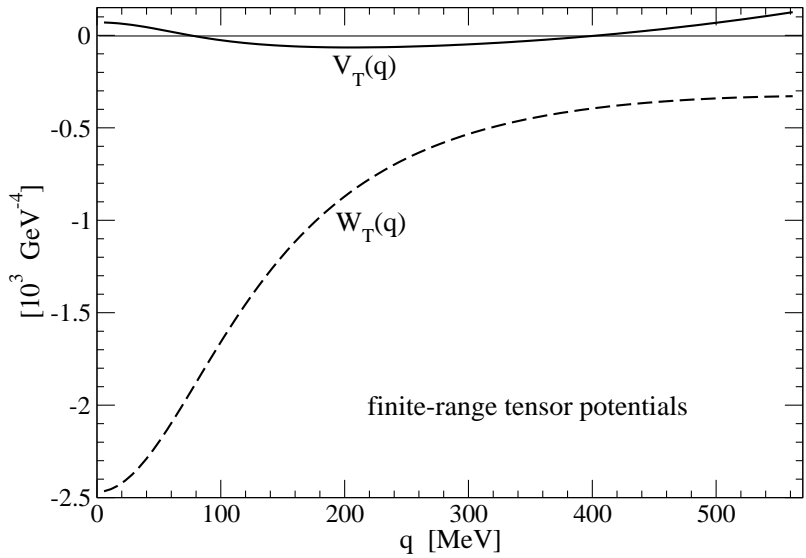}
\end{center}
\vspace{-.8cm}
\caption{Finite-range isoscalar and isovector tensor potentials extracted from 
N$^3$LOW \cite{n3low}.}
\end{figure}

\begin{figure}
\begin{center}
\includegraphics[scale=0.43,clip]{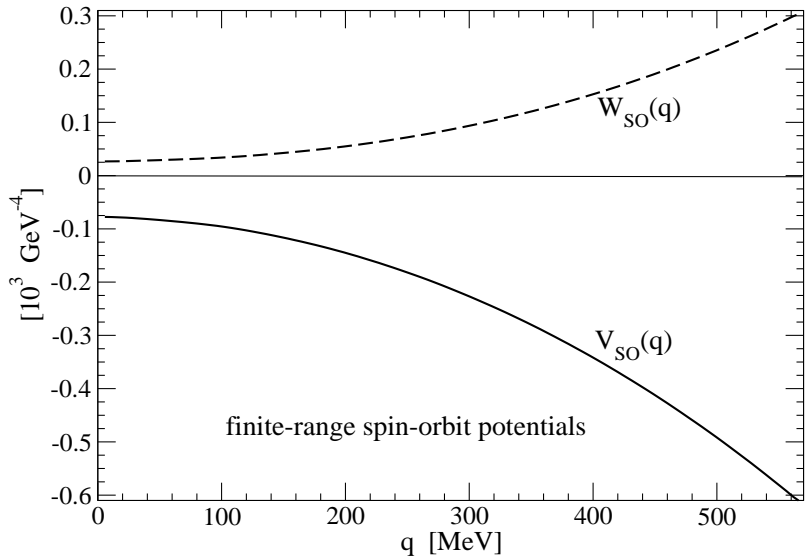}
\end{center}
\vspace{-.8cm}
\caption{Finite-range isoscalar and isovector spin-orbit potentials extracted from 
N$^3$LOW \cite{n3low}.}
\end{figure}

The solid and dashed lines in Figs.\,1,2,3,4 show the finite-range isoscalar and isovector 
potentials extracted from the chiral NN-interaction N$^3$LOW  \cite{n3low} in the central, 
spin-spin, tensor, and spin-orbit channel, respectively. In each figure the curves extend 
up to momentum transfers of $q= 570\,$MeV, corresponding to the region $q< 2k_f$ within 
which the interaction gets probed for nuclear densities up to $\rho = 0.2\,$fm$^{-3}$. One 
notices in Fig.\,3 the large negative values of $W_T(q)$ (multiplied with $q^2/3$ in 
$V_{NN}^{(\pi)}$) which result at small momentum transfers from $1\pi$-exchange.

In the (first-order) Hartree-Fock approximation the finite-range NN-potential 
$V_{NN}^{(\pi)}$ leads in combination with the density matrix-expansion (i.e. by employing 
the product of two medium insertions $\Gamma(\vec p_1,\vec q\,)\,\Gamma(\vec p_2,
-\vec q\,)$), to the following two-body contributions to the energy density functional 
${\cal E}[\rho,\tau,\vec J\,]$:  
\begin{eqnarray} \bar E(\rho) &=& {\rho\over 2} V_C(0) - {3\rho\over 2} \int_0^1
\!\!dx\, x^2(1-x)^2(2+x) \Big[V_C(q)+3W_C(q) \nonumber \\ && \qquad\qquad\quad
+3V_S(q)+9W_S(q) +q^2V_T(q) +3q^2W_T(q)\Big] \,, \end{eqnarray}
\begin{equation} F_\tau(\rho) = {k_f\over 2\pi^2}\int_0^1\!\!dx(x-2x^3)\Big[
V_C(q)+3W_C(q) +3V_S(q)+9W_S(q) +q^2V_T(q) +3q^2W_T(q)\Big] \,, \end{equation}
\begin{equation} F_d(\rho) = {1\over 4} V_C''(0)\simeq -27.1\,{\rm MeV fm}^5 \,, 
\end{equation} 
\begin{equation} F_{so}(\rho) = {1\over 2}V_{SO}(0)+ \int_0^1\!\!dx\,x^3\Big[
V_{SO}(2x k_f)+3W_{SO}(2x k_f)\Big] \,, \end{equation}
\begin{equation} F_J(\rho) = {3\over 8k_f^2} \int_0^1\!\!dx\Big\{(2x^3-x)\Big[
V_C(q)+3W_C(q) -V_S(q)-3W_S(q)\Big] -x^3 q^2\Big[V_T(q) +3W_T(q)\Big]\Big\}
\,, \end{equation} 
setting $q = 2x k_f$. The double-prime in eq.(9) denotes a second derivative and we 
have given the numerical value  for $F_d(\rho)$ resulting from the negative curvature of 
isoscalar central potential $V_C(q)$ shown in Fig.\,1. One can easily convince oneself 
that $F_J(\rho)$ as given in eq.(11) stays finite in the limit $k_f \to 0$. After 
expanding the integrand to linear order in $q^2$, the constant $V_C(0)+3W_C(0)$ 
integrates to zero. 

In addition to the finite-range pieces written in eqs.(7-11) there are the two-body 
contributions from the zero-range contact potential of the chiral NN-interaction 
N$^3$LOW. The  corresponding expression in momentum space includes constant, quadratic, 
and quartic terms in momenta and it can be found in section 2.2 of ref.\cite{evgeni}. 
The Hartree-Fock contributions from the NN-contact potential to the nuclear energy 
density functional ${\cal E}[\rho,\tau,\vec J\,]$ read: 
\begin{equation} \bar E(\rho) = {3\rho\over 8}(C_S-C_T)+{3 \rho k_f^2 \over 20} 
(C_2-C_1-3C_3-C_6)+{9\rho k_f^4\over 140}(D_2-4D_1-12D_5-4D_{11})\,,\end{equation} 
\begin{equation} F_\tau(\rho) = {\rho \over 4} (C_2-C_1-3C_3-C_6)+{\rho k_f^2\over
 4}(D_2-4D_1-12D_5-4D_{11})\,,\end{equation} 
\begin{equation} F_d(\rho) = {1\over 32} (16C_1-C_2-3C_4-C_7)+{k_f^2\over 48}
(9D_3+6D_4-9D_7-6D_8-3D_{12}-3D_{13}-2D_{15})\,,\end{equation} 
\begin{equation} F_{so}(\rho) = {3\over 8} C_5+{k_f^2\over 6}(2D_9+D_{10})\,,
\end{equation}
\begin{equation} F_J(\rho) = {1\over 16} (2C_1-2C_3-2C_4-4C_6+C_7)+{k_f^2\over 32}
(16D_1-16D_5-4D_6-24D_{11}+D_{14})\,.\end{equation} 
The 24 low-energy constants $C_{S,T}$, $C_j$ and $D_j$ are determined (at the cut-off 
scale of $\Lambda = 414\,$MeV) in fits to empirical NN-phase shifts and deuteron 
properties \cite{n3low}. We have extracted their values from the pertinent NN-scattering 
code made available to us by R. Machleidt. The values of the low-energy constants for the 
pure contact terms are:  $C_S=-117.5$, $C_T=2.937$ (in units GeV$^{-2}$), those of the terms 
quadratic in momenta are: $C_1 = 475.8$, $C_2 = 1034.1$, $C_3 = -29.04$, $C_4 = -524.4$, 
$C_5 = 717.4$, $C_6 = -42.70$, $C_7 = -1753.6$ (in units GeV$^{-4}$), and those of the terms 
quartic in momenta are: $D_1 = 0.612$, $D_2 = 25.61$, $D_3= 19.68$, $D_4= -19.49$, 
$D_5= 1.287$, $D_6=19.02$, $D_7 = 6.565$, $D_8 = -5.429$, $D_9 = 4.226$, $D_{10} = 
-16.02$, $D_{11} = -1.243$, $D_{12} = -0.976$, $D_{13} = -0.998$, $D_{14} = -7.995$, 
$D_{15} = -0.491$ (in units $10^3\,$GeV$^{-6}$). Let us mention that the 
contributions proportional to $C_{S,T}$ and $C_j$ in eqs.(12-16) have also been worked 
out in appendix B of ref.\cite{microefun} and we find agreement with their results. The 
terms proportional to $D_j$ as well as the master formulas eqs.(7-11) 
for the finite-range contributions are new. Note that we do not include an additional 
regulator function \cite{n3low} since the NN-interactions are probed only at small 
momenta $|\vec p_{1,2}|\leq k_f\leq 285\,$MeV.     
\section{Three-body contributions}
In this section we work out the three-body contributions to the nuclear energy density 
functional ${\cal E}[\rho,\tau,\vec J\,]$. We employ the leading order chiral 
three-nucleon interaction \cite{3bodycalc} which consists of a contact piece 
(with parameter $c_E$), a $1\pi$-exchange component (with parameter $c_D$) and a 
$2\pi$-exchange component (with parameters $c_1$, $c_3$ and $c_4$). In order to treat 
the three-body correlations in inhomogeneous nuclear many-body systems we follow 
ref.\cite{efun} and assume that the relevant product of density-matrices can be 
represented in momentum space in a factorized form by $\Gamma(\vec p_1,\vec q_1)\,
\Gamma(\vec p_2,\vec q_2)\,\Gamma(\vec p_3,-\vec q_1-\vec q_2)$. Such a factorization 
ansatz respects by construction the correct nuclear matter limit, but it involves 
approximations in comparison to more sophisticated treatments outlined in section 4 of 
ref.\cite{platter}. Actually, our approach is similar to the method DME-I introduced in
ref.\cite{platter}.

\begin{figure}
\begin{center}
\includegraphics[scale=1.2,clip]{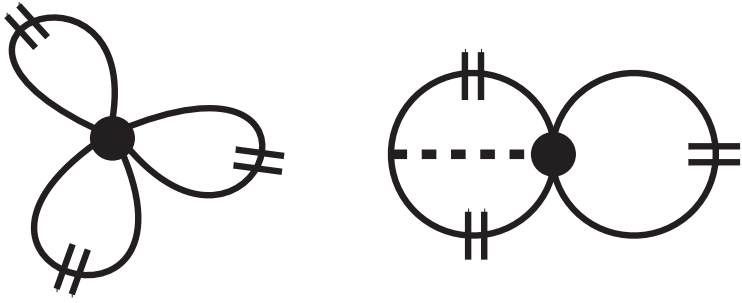}
\end{center}
\vspace{-.6cm}
\caption{Three-body diagrams related to the contact ($c_E$) and $1\pi$-exchange ($c_D$) 
component of the chiral three-nucleon interaction. The short double-line symbolizes the 
medium insertion $\Gamma(\vec p, \vec q\,)$ for inhomogeneous nuclear matter.}
\end{figure}

\subsection{$c_E$-term}
We start with the three-body contributions from the contact interaction as
represented by the left diagram in Fig.\,5. With three (inhomogeneous) medium insertions 
one finds the following  contribution to the energy per particle:
\begin{equation} \bar E(\rho) = -{c_E k_f^6 \over 12 \pi^4 f_\pi^4\Lambda_\chi}\,,
\end{equation}
which is quadratic in the density $\rho=2k_f^3/3\pi^2$. Obviously, the contributions to 
the other strength functions $F_{\tau,d,so,J}(\rho)$ vanish due to the 
momentum-independence of the contact interaction. For the choice of scale $\Lambda_{\chi}
= 700$\,MeV the value $c_E = -0.625$ has been determined in calculations of few-nucleon 
systems \cite{achim} (employing in addition $V_{\rm low-k}$ for the two-body 
interaction).

\subsection{$c_D$-term}
Next, we consider the three-body contributions from the $1\pi$-exchange component of the 
chiral 3N-interaction as represented by the right diagram in Fig.\,5. Putting in three 
(inhomogeneous) medium insertions one finds the following analytical expressions:  
\begin{equation} \bar E(\rho)={g_A c_D m_\pi^6\over(2\pi f_\pi)^4\Lambda_\chi}
\bigg\{{u^6 \over3}-{3u^4\over 4}+{u^2\over 8} +u^3 \arctan 2u -{1+12u^2 \over
  32}\ln(1+4u^2) \bigg\}\,,\end{equation}  
\begin{equation} F_\tau(\rho)={2g_A c_D m_\pi^4\over(4\pi f_\pi)^4\Lambda_\chi}
\Big\{(1+2u^2)\ln(1+4u^2)-4u^2 \Big\}\,,\end{equation}
\begin{equation} F_d(\rho)={g_A c_D m_\pi\over(4f_\pi)^4 \pi^2\Lambda_\chi}
\bigg\{{1\over 2u}\ln(1+4u^2)-{2u\over 1+4u^2} \bigg\}\,,\end{equation}
\begin{equation} F_J(\rho)={3g_A c_D m_\pi\over(4f_\pi)^4 \pi^2\Lambda_\chi}
\bigg\{ 2u-{1\over u}+{1\over 4u^3}\ln(1+4u^2)\bigg\}\,, \end{equation}
with the abbreviation $u=k_f/m_\pi$. For the parameter $c_D$ we take the value 
$c_D = -2.06$ from ref.\cite{achim}. Note that there is no contribution to 
the spin-orbit coupling strength $F_{so}(\rho)$. 
 \subsection{Hartree diagram proportional to $c_{1,3}$}
\begin{figure}
\begin{center}
\includegraphics[scale=1.2,clip]{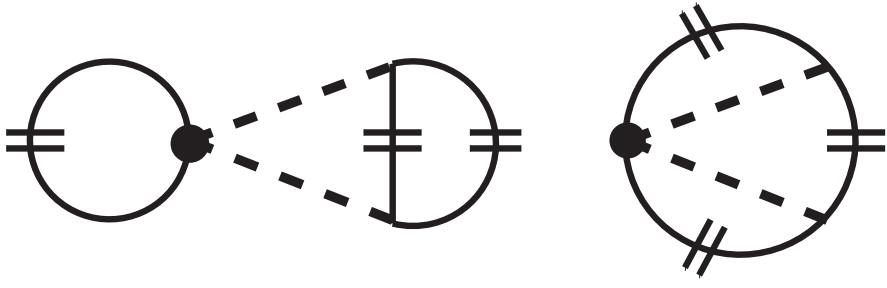}
\end{center}
\vspace{-.6cm}
\caption{Three-body Hartree and Fock diagrams related to the chiral $2\pi$-exchange 
three-nucleon interaction.}
\end{figure}

We continue with the three-body contributions from the $2\pi$-exchange Hartree diagram 
shown in the left part of Fig.\,6. Again with three (inhomogeneous) medium insertions one 
derives the following analytical results:
\begin{eqnarray}  \bar E(\rho)&=&{g_A^2 m_\pi^6\over(2\pi f_\pi)^4}\bigg\{(12c_1
-10c_3) u^3\arctan 2u -{4\over 3} c_3 u^6 +6(c_3-c_1)u^4 \nonumber \\ && +(3c_1
-2c_3)u^2 +\bigg[{1\over 4}(2c_3-3c_1)+{3u^2\over 2}(3c_3-4c_1)\bigg]\ln(1+4u^2)
\bigg\}\,,  \end{eqnarray}
\begin{eqnarray}  F_\tau(\rho)&=&{g_A^2 m_\pi^4\over(2\pi   f_\pi)^4}\bigg\{
(5c_3-6c_1)u^2 +{(c_3-2c_1)u^2 \over 1+4u^2}\nonumber \\ && +\bigg[
2c_1-{3\over 2}c_3+2(c_1-c_3)u^2\bigg]\ln(1+4u^2) \bigg\}\,,  \end{eqnarray}
\begin{eqnarray}  F_d(\rho)&=&{g_A^2 m_\pi\over  (8\pi)^2 f_\pi^4}\bigg\{(10c_1-
23c_3)\arctan 2u+ 16c_3 u  \nonumber \\ && +{7c_3-5c_1 \over u}\ln(1+4u^2) 
+ {6c_3 u+16(2c_3-c_1)u^3\over 3(1+4u^2)^2} \bigg\}\,,  \end{eqnarray}
\begin{equation} F_{so}(\rho)  = {3g_A^2 m_\pi \over (8\pi)^2 f_\pi^4}
\bigg\{{2\over u}(4c_1-3c_3) -4c_3 u+\bigg[{4\over u}(c_3-c_1)+{3c_3-4c_1
    \over 2u^3}\bigg] \ln(1+4u^2) \bigg\}\,, \end{equation}
\begin{equation} F_J(\rho)  = {3g_A^2 m_\pi \over (8\pi)^2 f_\pi^4}\bigg\{
{3c_3-4c_1 \over u} -2c_3 u+{4u(2c_1-c_3)\over 1+4u^2}+{4c_1-3c_3 \over 4u^3}
\ln(1+4u^2) \bigg\}\,, \end{equation}
which depend only on the two isoscalar coupling constants $c_1$ and $c_3$ with values   
$c_1=-0.76$\,GeV$^{-1}$ and $c_3=-4.78$\,GeV$^{-1}$ \cite{3bodycalc}. Note that the 
expression for $F_{so}(\rho)$ in eq.(25) gives the dominant part of the three-body 
spin-orbit coupling strength suggested originally by Fujita and Miyazawa \cite{fujita}.
Their proposed mechanism is based on the excitation of a $\Delta(1232)$-resonance.
In the present approach the two-step process $\pi N\to \Delta \to \pi N$ is replaced by 
an equivalent $\pi\pi NN$ contact vertex proportional to $c_3$. 

\subsection{Fock diagram proportional to $c_{1,3,4}$}
Finally, there are the three-body contributions from the $2\pi$-exchange Fock diagram 
shown in the right part of Fig.\,6. With one single closed nucleon ring this diagram 
generates (for isospin-symmetric nuclear matter) also non-vanishing contributions from 
the isovector $\pi\pi NN$ contact vertex proportional to $c_4$. We take consistently the 
value $c_4 =3.96$\,GeV$^{-1}$ used in few-body calculations by ref.\cite{3bodycalc}. 
In the case of the three-body Fock diagram not all of the occurring integrals over the 
three Fermi spheres can be solved analytically. Collecting all the emerging pieces, we 
find the following results for the Fock contributions to the nuclear energy density 
functional ${\cal E}[\rho,\tau,\vec J\,]$:
\begin{eqnarray} \bar E(\rho)&=&{3g_A^2 m_\pi^6\over(4\pi f_\pi)^4 u^3}
\int_0^u\!\!dx \bigg\{\bigg({c_3 \over 2}-c_4\bigg) G_S^2 +(c_3+c_4)G_T^2+ 3c_1
\nonumber \\ && \times\Big[u(1+u^2+x^2)-\Big(1+(u+x)^2\Big)\Big(1+(u-x)^2\Big)
L\Big]^2 \bigg\}\,, \end{eqnarray}
\begin{eqnarray} F_\tau(\rho)&=&{g_A^2 m_\pi^4\over(4\pi f_\pi)^4}\Bigg\{ {2\over 
u^3}(2c_4-c_3)\Big[4u^2-(1+2u^2)\ln(1+4u^2) \Big]\arctan 2u\nonumber  \\ && +
{1+2u^2 \over 64 u^8}\Big[48(c_4-c_3-2c_1)u^4-8(3c_1+2c_3+2c_4)u^2-3(c_3+c_4)\Big] 
\nonumber \\ &&\times \ln^2(1+4u^2) +\bigg[{4\over 3}(c_3-5c_4)u^2+6c_1-{7\over 3}
c_3+{17\over 3}c_4+{3\over 8u^6}(c_3+c_4)\nonumber  \\ && +{24c_1+5(c_3+c_4)\over
 2u^2}+{3c_1+2(c_3+c_4) \over u^4}\bigg] \ln(1+4u^2) +{8\over 3}(c_3+c_4)u^4 
\nonumber  \\ && +{2u^2 -3\over 3}(17c_4-7c_3)-12c_1-{12c_1+5(c_3+c_4) 
\over 2u^2} -{3\over 4u^4}(c_3+c_4)\nonumber  \\ && +{1\over u^3} \int_0^u\!\!dx
\bigg\{12c_1 \Big[u(1+u^2+x^2)-\Big(1+(u+x)^2\Big)\Big(1+(u-x)^2\Big)L\Big]
\nonumber \\ &&\times  \bigg[(1-u^2-x^2) L +u -{u \over 1+(u+x)^2} -{u \over 
1+(u-x)^2}\bigg]\nonumber \\ && +(2c_4-c_3)G_S\bigg[{2u(u+x)\over 1+(u+x)^2}+
{2u(x-u)\over 1+(u-x)^2} - 4 x L\bigg]\nonumber \\ && -(c_3+c_4) G_T \bigg[{3u 
\over x}(3u^2-1)-3u x+{4u(u+x)\over 1+(u+x)^2}+{4u(x-u) \over 1+(u-x)^2} 
\nonumber \\ && +{L\over x}(3x^4+6u^2x^2-2x^2-9u^4-6u^2+3) \bigg] \bigg\}\Bigg\} 
\,, \end{eqnarray}
\begin{eqnarray} F_d(\rho)&=&{g_A^2 m_\pi\over \pi^2(4f_\pi)^4}\Bigg\{  c_1\bigg[
{16u \over 1+4u^2} -{6\over u}+\bigg({3\over u^3}-{4\over u}+{16u \over  1+4u^2}
\bigg) \ln(1+4u^2)\nonumber \\ && -{3\over 8u^5} \ln^2(1+4u^2)\bigg] +(c_3+c_4)
\bigg[{3\over u}-{3\over 2u^3}-{12u \over 1+4u^2}\nonumber \\ &&
+\bigg({3\over 4u^5}+{5\over 2u}-{8u \over 1+4u^2}\bigg) \ln(1+4u^2)-{3+6u^2+
8u^4 \over 32u^7} \ln^2(1+4u^2)\bigg] \nonumber \\ && +c_4 \bigg[{8u \over 1+4u^2}
+\bigg({8u \over 1+4u^2}-{4\over u}\bigg)\ln(1+4u^2)+{1\over 2u^3}\ln^2(1+4u^2)
\bigg] \Bigg\} \,,\end{eqnarray}
\begin{eqnarray}F_{so}(\rho) &=& {g_A^2 m_\pi \over \pi^2 (4 f_\pi u)^4}\Bigg\{ 
3c_1\bigg[ 2u-2u^3+{3\over 2u} -{3+10u^2 \over 4u^3} \ln(1+4u^2) \nonumber \\
&& +{3+16u^2+16u^4 \over 32 u^5} \ln^2(1+4u^2)\bigg] + (c_3+c_4) \bigg[u^3-{16 
u^5\over 3} +{7u \over 4}\nonumber \\ && +{3 \over u}+{15\over 16u^3}+\bigg(
2u^3-{3u\over 2}-{13\over 4u}-{39 \over 16u^3}-{15\over  32u^5}\bigg)\ln(1+4u^2)
\nonumber \\ && +{3\over 256 u^7}(64u^6+80u^4+36u^2+5)  \ln^2(1+4u^2)\bigg]
\Bigg\}\,,\end{eqnarray}
\begin{eqnarray}F_J(\rho) &=& {9g_A^2 c_1 m_\pi \over \pi^2 (4 f_\pi u)^4}\Bigg\{ 
{10 u^3 \over 3}+{11u\over 8}-{1\over 2u} + {8+9u^2-76u^4 \over 32 u^3}
\ln(1+4u^2) \nonumber \\ && -{1\over 2}\arctan 2u -{1+4u^2 \over 32 u^5}\ln^2
(1+4u^2) + \int_0^u\!\!dx\bigg\{ {L^2 \over u^2}\bigg[{3\over 4x^2} (1+u^2)^4
\nonumber \\ && +(1+u^2)(1-u^4)+{11 x^6\over 4} +5(1-u^2)x^4 +{x^2\over 2} 
(5u^4-14u^2+5)\bigg]\nonumber \\ && +{L\over 2u} \bigg[3u^4+2u^2-1-{3\over  x^2}
(1+u^2)^3\bigg] +{3\over 4x^2}(1+u^2)^2 \bigg\} \Bigg\} \nonumber \\ &+&
{3g_A^2 c_3 m_\pi \over \pi^2 (8 f_\pi u)^4}\Bigg\{ \Big[7+65u^2-34u^4+8u^{-2}  
\ln(1+4u^2)\Big]\arctan 2u \nonumber \\ && +{832u^5 \over 5}-{1415u^3\over 12}
+{91u\over 4}-{4\over u}-{3\over u^3} -{3+16u^2+48u^4 \over 16 u^7} \nonumber
\\ &&  \times  \ln^2(1+4u^2) + \bigg( {3\over  2u^5} +{5\over u^3}-{103 \over 
16u}+{221u \over 48}-{15u^3 \over  4}\bigg) \ln(1+4u^2)\nonumber \\ && 
+\int_0^u\!\!dx\,\bigg\{{3L^2 \over 2u^2}\bigg[{5\over x^4}(1+u^2)^6 +{6\over x^2}
(1+u^2)^4(1-3u^2) +(1+u^2)^2 \nonumber \\ && \times (23-18u^2+39u^4) + 4x^2(9+23
u^2-5u^4-19u^6) +17x^8 \nonumber \\ && +x^4(19-26u^2+99u^4) +22x^6(1-3u^2)\bigg]
+{L \over u}\bigg[-{15\over  x^4}(1+u^2)^5\nonumber \\ && +{1\over  x^2}(1+u^2)^3
(49u^2-3) -6(17u^6+13u^4+7u^2+11) \bigg] \nonumber \\ && +{15\over  2x^4}
(1+u^2)^4 -{2\over  x^2}(1+u^2)^2(3+11u^2) \bigg\}\Bigg\}\nonumber \\ &+&
{3g_A^2 c_4 m_\pi \over \pi^2 (8 f_\pi u)^4}\Bigg\{ \Big[79-95u^2-10u^4-16u^{-2}  
\ln(1+4u^2)\Big]\arctan 2u \nonumber \\ && -{512u^5 \over 15}+{2185u^3\over 12}
-{181u\over 4}-{4\over u}-{3\over u^3} +{48u^4-16u^2-3 \over 16 u^7} \nonumber
\\ &&  \times  \ln^2(1+4u^2) + \bigg( {3\over  2u^5} +{5\over u^3}-{119 \over 
16u}+{173u \over 48}+{9u^3 \over  4}\bigg) \ln(1+4u^2)\nonumber \\ && 
+\int_0^u\!\!dx\,\bigg\{{3L^2 \over 2u^2}\bigg[{5\over x^4}(1+u^2)^6 +{6\over x^2}
(1+u^2)^4(1-3u^2) +(1+u^2)^2 \nonumber \\ && \times (7+14u^2+23u^4) + 4x^2(9+7
u^2-5u^4-3u^6) +x^8 \nonumber \\ && +x^4(51-26u^2+3u^4) +x^6(22-2u^2)\bigg]
+{L \over u}\bigg[-{15\over  x^4}(1+u^2)^5\nonumber \\ && +{1\over  x^2}(1+u^2)^3
(49u^2-3) -18(1+3u^2)(1+u^2)^2 \bigg] \nonumber \\ && +{15\over  2x^4}(1+u^2)^4 
-{2\over  x^2}(1+u^2)^2(3+11u^2) \bigg\}\Bigg\}
\,.\end{eqnarray}
Here we have introduced the auxiliary functions: 
\begin{eqnarray} G_S(x,u) &=& {4ux \over 3}( 2u^2-3) +4x\Big[
\arctan(u+x)+\arctan(u-x)\Big] \nonumber \\ && + (x^2-u^2-1) \ln{1+(u+x)^2
\over  1+(u-x)^2} \,,\\ G_T(x,u) &=& {ux\over 6}(8u^2+3x^2)-{u\over
2x} (1+u^2)^2  \nonumber \\ && + {1\over 8} \bigg[ {(1+u^2)^3 \over x^2} -x^4 
+(1-3u^2)(1+u^2-x^2)\bigg] \ln{1+(u+x)^2\over  1+(u-x)^2} \,,\\
 L(x,u) &=& {1\over 4x}\ln{1+(u+x)^2\over  1+(u-x)^2} \,,\end{eqnarray}
where $u=k_f/m_\pi$. A good check of all formulas collected in this section is provided by 
their Taylor expansion in $k_f$. Despite the superficial first appearance to the contrary, 
one can verify that the leading term in the $k_f$-expansion is $k_f^3$. In several cases 
it is even a higher power of $k_f$.
\section{Results and discussion}
\begin{figure}
\begin{center}
\includegraphics[scale=0.43,clip]{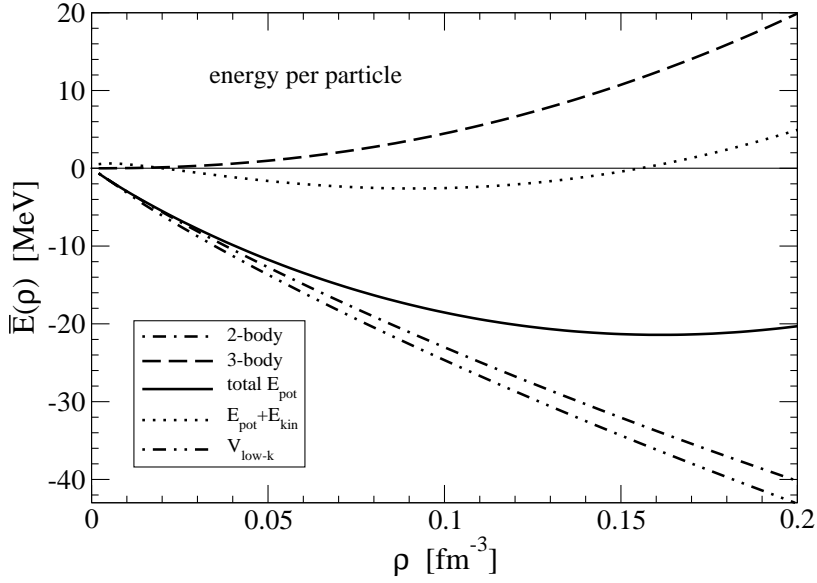}
\end{center}
\vspace{-.8cm}
\caption{Contributions to the energy per particle $\bar E(\rho)$ of isospin-symmetric 
nuclear matter.}
\end{figure}

In this section we present and discuss our numerical results obtained by summing the 
series of two- and three-body contributions given in sections 3 and 4. The physical 
input parameters are: $g_A=1.3$ (nucleon axial vector coupling constant), 
$f_\pi= 92.4\,$MeV (pion decay constant) and $m_\pi=138\,$MeV (average pion mass). The 
parameters $C_{S,T}$, $C_j$ and $D_j$ corresponding to the 24 contact terms in the chiral 
NN-potential N$^3$LOW \cite{n3low} have already been listed at the end of section 3. For 
the sake of completeness we quote again the parameters pertinent to the chiral 
three-nucleon interaction: $c_E=-0.625$, $c_D=-2.06$, $\Lambda_\chi=700\,$MeV, $c_1=
-0.76\,$GeV$^{-1}$, $c_3=-4.78\,$GeV$^{-1}$ and $c_4=3.96\,$GeV$^{-1}$, taken from 
refs.\cite{achim,3bodycalc}.

Fig.7 shows the contributions to the energy per particle $\bar E(\rho)$ of 
isospin-symmetric nuclear matter for densities up to $\rho = 0.2\,$fm$^{-3}$. The 
dash-dotted line gives the (attractive) two-body contributions and the dashed line the 
(repulsive) three-body contributions. For comparison we have also included the 
Hartree-Fock contribution to the energy per particle $\bar E(\rho)$ as obtained from the 
universal low-momentum NN-potential $V_{\rm low-k}$ \cite{vlowkreview,vlowk,achim} by summing 
and integrating its diagonal (on-shell) partial-wave matrix elements. One observes that 
our treatment of the NN-interaction via the chiral potential N$^3$LOW reproduces these 
results fairly accurately. The sum of the two- and three-body contributions (full line in 
Fig.\,7) shows a first tendency for saturation of nuclear matter. However, after 
inclusion of the kinetic energy $\bar E_{\rm kin}(\rho) = 3k_f^2/10M-3k_f^4/56M^3$ the 
resulting minimum is still much too shallow. This observation (at the Hartree-Fock level)
is consistent with refs.\cite{achim,hebeler}. An improved description of the nuclear 
matter equation of state $\bar E(\rho)$ can be achieved when treating the 
two-body interaction at least to second order.

\begin{figure}
\begin{center}
\includegraphics[scale=0.43,clip]{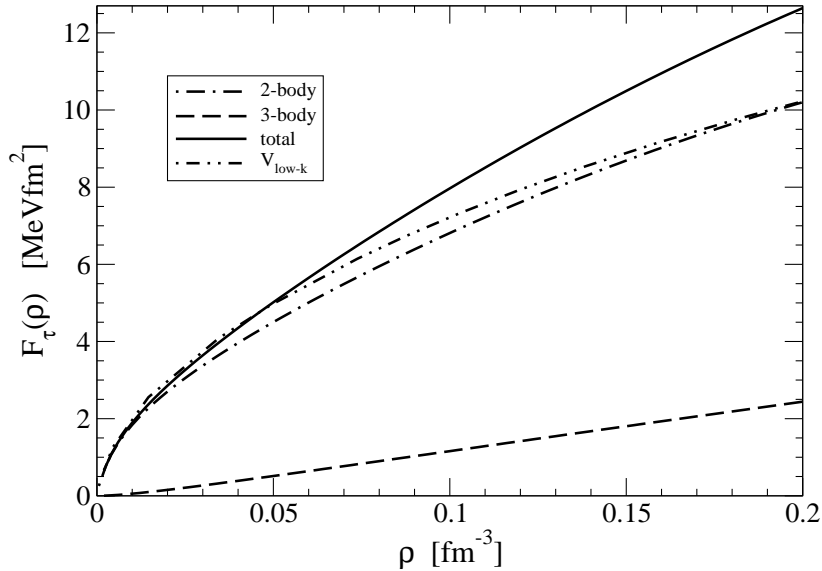}
\end{center}
\vspace{-.6cm}
\caption{Contributions to the strength function $F_\tau(\rho)$ versus the nuclear 
density $\rho$.}
\end{figure}
\begin{figure}
\begin{center}
\includegraphics[scale=0.43,clip]{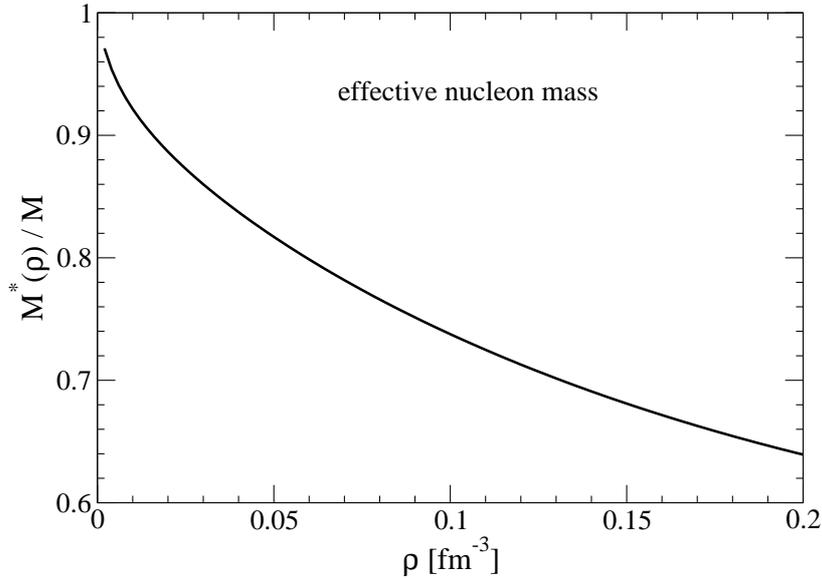}
\end{center}
\vspace{-.8cm}
\caption{Effective nucleon mass $M^*(\rho)$ divided by free nucleon mass $M$ as 
function of $\rho$.}
\end{figure}

Fig.\,8 shows the contributions to the strength function $F_\tau(\rho)$. For the two-body 
part the results derived with $V_{\rm low-k}$ and the chiral N$^3$LOW potential lie
closely together. The three-body part (shown by the dashed line in Fig.\,8) comes out 
relatively small. At  $\rho_0=0.16\,$fm$^{-3}$ it adds a correction of about 
$20\%$. The expression multiplying the kinetic energy density $\tau(\vec r\,)$ in the 
nuclear energy density functional eq.(3) has the meaning of a reciprocal 
density-dependent effective nucleon mass:
\begin{equation} M^*(\rho) = M\bigg[ 1 -{k_f^2 \over 2M^2}+2M\, F_\tau(\rho)\bigg]^{-1}
\,. \end{equation}
It is identical to the so-called ''Landau''-mass introduced in Fermi-liquid theory, since 
it derives in the same way from the slope of the single-particle potential $U(p,k_f)$ at 
the Fermi surface $p = k_f$. The (small) correction term $-k_f^2/2M^2$ accounts for
the relativistic increase of mass. Fig.\,9 shows the ratio of effective to free nucleon 
mass $M^*(\rho)/M$ as a function of the nuclear density $\rho$.  One observes a reduced 
effective nucleon mass which reaches the value  $M^*(\rho_0) \simeq 0.67M$ at nuclear 
matter saturation density $\rho_0=0.16\,$fm$^{-3}$. This is compatible with the range 
$0.7 < M^*(\rho_0)/M<1$ spanned by phenomenological Skyrme forces \cite{sly,pearson}. On 
the other hand it has been found recently in ref.\cite{landau2nd} that second-order 
corrections from $V_{\rm low-k}$ enhance the effective nucleon mass substantially.

\begin{figure}
\begin{center}
\includegraphics[scale=0.43,clip]{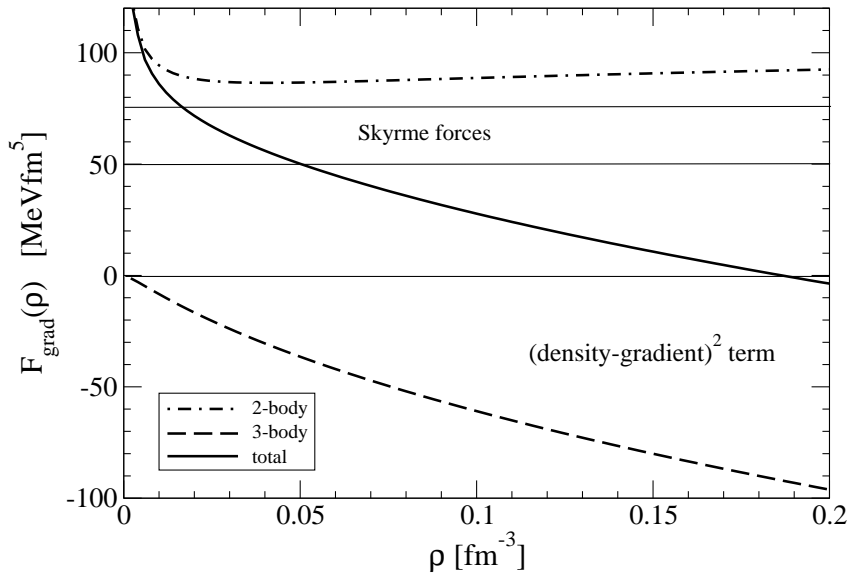}
\end{center}
\vspace{-.8cm}
\caption{The strength function $F_\nabla(\rho)$ of the surface term 
$(\vec \nabla \rho)^2$ versus the nuclear density $\rho$.}
\end{figure}
     
\begin{figure}
\begin{center}
\includegraphics[scale=0.43,clip]{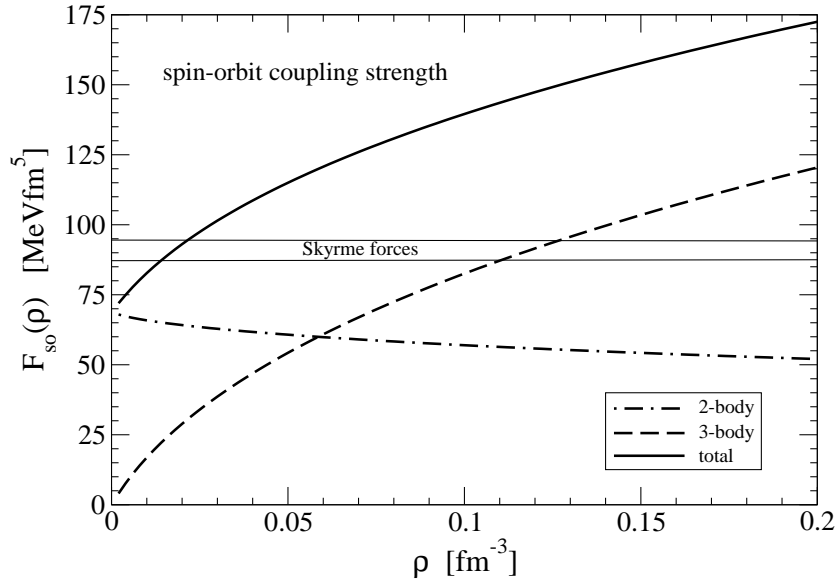}
\end{center}
\vspace{-.8cm}
\caption{Strength function $F_{so}(\rho)$ of the spin-orbit coupling term 
$\vec \nabla \rho\cdot \vec J$ versus $\rho$.}
\end{figure}

Next, we show in Fig.\,10 the strength function $F_\nabla(\rho)$ of the $(\vec 
\nabla \rho)^2$ surface-term. The pronounced increase of the two-body contribution 
(dash-dotted line) at very low densities is caused by the $1\pi$-exchange and has also 
been observed in other calculations \cite{efun,microefun}. The sizeable three-body 
contribution (dashed line) adds negatively to these initial values such that the total 
result for $F_\nabla(\rho)$ (shown by the full line in Fig.\,10) decreases with increasing 
density $\rho$. For comparison we have also included the band (of constant $F_\nabla
(\rho)$-values) spanned by phenomenological Skyrme forces  \cite{sly,pearson}. Taking 
this as a benchmark one sees that our Hartree-Fock result is somewhat too small at 
densities around $\rho_0/2 =0.08\,$fm$^{-3}$, where the surface energy in nuclei gains 
most of its weight. The calculation of the iterated $1\pi$-exchange in ref.\cite{efun} 
suggests that a treatment of the low-momentum two-body interaction to second order will 
further increase the values of $F_\nabla(\rho)$.

Of particular interest is the strength function $F_{so}(\rho)$ of the spin-orbit 
coupling term $\vec \nabla \rho\cdot \vec J$. The two- and three-body contributions 
together with their total sum are shown in Fig.\,11. The two-body part is dominated by 
the low-energy constant $3C_5/8$ as indicated also by the weak variation of the 
dash-dotted line with density $\rho$. The induced (density-dependent) three-body 
spin-orbit forces add sizeably to this initial value. The major contribution is provided 
by the Hartree term in eq.(23) proportional to $c_3=-4.78\,$GeV$^{-1}$. With this given 
value of $c_3$ it is considerably larger than the $\Delta(1232)$-excitation mechanism 
proposed by Fujita and Miyazawa \cite{fujita} which corresponds to $c_3^{(\Delta)} = -g_A^2/2
\Delta\simeq-2.9\,$GeV$^{-1}$ (with $\Delta = 293\,$MeV the delta-nucleon mass splitting).  
At densities around $\rho_0/2 = 0.08\,$fm$^{-3}$ where the spin-orbit interaction in nuclei 
receives most of its weight, our total Hartree-Fock result overshoots the empirical 
spin-orbit coupling strength $F_{so}^{\rm (emp)}(\rho) \simeq 90\,$MeVfm$^{5}$ \cite{sly,
pearson} by about $50\%$. This requires a compensating effect, and indeed it has been 
found in  ref.\cite{efun} that the second-order $1\pi$-exchange tensor force generates a 
spin-orbit coupling of the ''wrong-sign''. Taking the value $F_{so}^{\rm(1\pi-it)}(\rho_0/2)
\simeq -35\,$MeVfm$^{5}$ (see Fig.\,5 in ref.\cite{efun}) as indicative one can expect 
that the second-order effects from the low-momentum two-nucleon tensor potential will 
reduce the strength of the spin-orbit coupling $F_{so}(\rho)$ to the correct amount.

\begin{figure}
\begin{center}
\includegraphics[scale=0.43,clip]{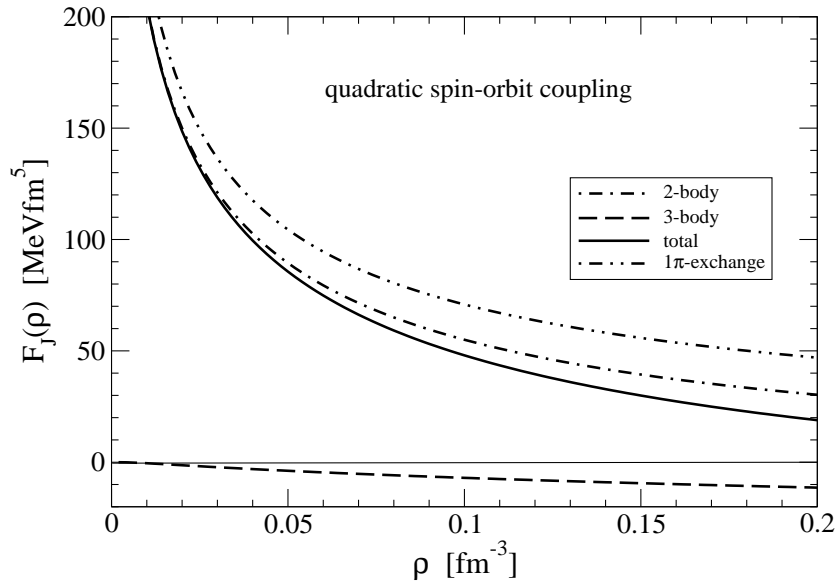}
\end{center}
\vspace{-.8cm}
\caption{The strength function $F_J(\rho)$ multiplying the squared spin-orbit density 
$\vec J^{\,2}$ versus $\rho$.}
\end{figure}

Finally, we show in Fig.\,12 the strength function $F_J(\rho)$ of the squared 
spin-orbit density $\vec J^{\,2}$ in the nuclear energy density functional ${\cal E}
[\rho,\tau,\vec J\,]$. In contrast to all previous quantities, $F_J(\rho)$ receives 
only a very small three-body contribution. On the other hand the two-body contribution is 
strongly density-dependent and it reaches quite large values at low densities. This 
prominent feature of $F_J(\rho)$ has also been observed in previous calculations 
\cite{efun,microefun}. Actually, the strong density dependence of  $F_J(\rho)$ originates 
from the dominant $1\pi$-exchange contribution which we reproduce separately in 
Fig.\,12 by the upper dash-dotted line (for an explicit formula  for $F_J^{(1\pi)}(\rho)$, 
see eq.(11) in ref.\cite{efun}). At this point it should be kept in mind that the 
$\vec J^{\,2}$-term in the nuclear energy density functional represents the non-local Fock 
contributions from tensor forces etc. An outstanding $1\pi$-exchange contribution to the 
strength function  $F_J(\rho)$ is therefore not surprising.

\section{Summary and outlook}
In this work we have used the (improved) density-matrix expansion of ref.\cite{dmeimprov} 
to calculate the nuclear energy density functional from chiral two- and three-nucleon 
interactions. We have employed the low-momentum NN-potential N$^3$LOW \cite{n3low} which 
is composed of long-range multi-pion exchanges and a set of short-distance contact 
terms. The coefficients of the latter have been determined in fits to empirical NN-phase 
shifts. The leading order chiral three-nucleon interaction has been taken with its 
parameters $c_E, c_D$ and $c_{1,3,4}$ fixed in calculations of nuclear few-body systems 
\cite{achim,3bodycalc}. With this input the nuclear energy density functional ${\cal E}[
\rho,\tau,\vec J\,]$ has been derived to first order in many-body perturbation theory, i.e. 
in the Hartree-Fock approximation. For the effective nucleon mass $M^*(\rho)$ and the 
strength functions $F_\nabla(\rho)$ and $F_{so}(\rho)$ of the surface and spin-orbit terms 
we have found (in the relevant density range) reasonable agreement with results of 
phenomenological Skyrme forces. However, as indicated in particular by the nuclear matter 
equation of state $\bar E(\rho)$, an improved description of the energy density functional 
requires at least the treatment of the two-nucleon interaction to second order in 
many-body perturbation theory. It is furthermore expected that tensor forces at 
second-order generate an additional ''wrong-sign'' spin-orbit coupling \cite{efun} which 
compensates part of the strong three-body contribution to $F_{so}(\rho)$.

Such a consistent second-order calculation of the nuclear energy density functional 
represents a challenge, since a satisfactory generalization of the density-matrix expansion
\cite{microefun} to that situation has not yet been formulated. At second order one must 
properly account for the presence of energy denominators which induce further spatial and 
temporal non-localities and possibly even an orbital dependence of the resulting energy 
density functional. In a first simplified approach one could follow ref.\cite{efun} and 
approximate the energy denominators by the kinetic energies of free (on-shell) nucleons, 
eventually improved by the inclusion of an effective nucleon mass $M^*(\rho)$. In order to 
keep the second-order calculation manageable and in order to see leading effects, one would 
also restrict the low-momentum two-nucleon interaction to some dominant components, such as 
one-pion exchange with a suitable regularization of its strong tensor force at short 
distances. Studies along this line are underway.

\section*{Acknowledgement}
We thank R. Machleidt for providing us the code for elastic NN-scattering with the chiral 
potential N$^3$LOW.


\begin{thebibliography}{99}
\bibitem{reinhard} M. Bender, P.H. Heenen and P.G. Reinhard, {\it Rev. Mod.
Phys.} {\bf  75} (2003) 121.\vs
\bibitem{stone} J.R. Stone  and P.G. Reinhard, {\it Prog. Part. Nucl. Phys.} {\bf  58} 
(2007) 587.\vs
\bibitem{sly} E. Chabanat et al., \textit{Nucl. Phys.} \textbf{A627} (1997) 710; 
\textbf{A635} (1998) 231; and refs. therein.\vs
\bibitem{pearson} N. Chamel, S. Goriely and J.M. Pearson, {\it  Nucl. Phys.} 
{\bf A812} (2008) 72.\vs
\bibitem{serot} B.D. Serot and J.D. Walecka, {\it Int. J. Mod. Phys.} {\bf E6} (1997) 
515.\vs
\bibitem{ring} T. Niksic, D. Vretenar and P. Ring, nucl-th/1102.4193; to be published in 
{\it Prog. Part. Nucl. Phys.} (2011).\vs 
\bibitem{lesinski} T. Lesinski, T. Duguet, K. Bennaceur and J. Meyer, 
\textit{Eur. Phys. J.} \textbf{A40} (2009) 121.\vs 
\bibitem{drut} J.E. Drut, R.J. Furnstahl and L. Platter, {\it Prog. Part. 
Nucl. Phys.} {\bf 64} (2010) 120; nucl-th/0906.1463.\vs
\bibitem{platter} S.K. Bogner, R.J. Furnstahl and L. Platter, \textit{Eur. 
Phys. J.} \textbf{A39} (2009) 219.\vs
\bibitem{drut2} J.E. Drut and L. Platter, nucl-th/1104.4357.\vs
\bibitem{vlowkreview} S.K. Bogner, R.J. Furnstahl and A. Schwenk, {\it Prog. Part.
Nucl. Phys.} {\bf 65} (2010) 94.\vs
\bibitem{vlowk}  S.K. Bogner, T.T.S. Kuo and A. Schwenk, {\it Phys. Reports} {\bf 386} 
(2003) 1.\vs
\bibitem{achim} S.K. Bogner, R.J. Furnstahl, A. Nogga and A. Schwenk, {\it Nucl. Phys.} 
{\bf A763} (2005) 59.\vs
\bibitem{hebeler} K. Hebeler et al, {\it Phys. Rev.} {\bf C83} (2011) 031301.\vs
\bibitem{roth} R. Roth et al.,  {\it Phys. Rev.} {\bf C73} (2006) 044312.\vs
\bibitem{negele} J.W. Negele and D. Vautherin, {\it Phys. Rev.} {\bf C5} (1972)
1472.\vs 
\bibitem{dmeimprov} B. Gebremariam, T. Duguet and S.K. Bogner, {\it Phys. Rev.} {\bf C82} 
(2010) 014305.\vs 
\bibitem{efun} N. Kaiser and W. Weise, \textit{Nucl. Phys.} \textbf{A836} (2010) 
256.\vs
\bibitem{microefun} B. Gebremariam, S.K. Bogner and T. Duguet, \textit{Nucl. 
Phys.} \textbf{A851} (2011) 17.\vs
\bibitem{stoitsov} M. Stoitsov et al., {\it Phys. Rev.} {\bf C82} (2010) 054307.\vs
\bibitem{mach} R. Machleidt and D.R. Entem, {\it Phys. Reports} {\bf 503} (2011) 1.\vs 
\bibitem{n3low} L. Coraggio et al., {\it Phys. Rev.} {\bf C75} (2007) 024311.\vs
\bibitem{3bodycalc}  A. Nogga,  S.K. Bogner  and A. Schwenk, {\it Phys. Rev.} 
{\bf C70} (2004) 061002.\vs
\bibitem{relcor} N. Kaiser,  {\it Phys. Rev.} {\bf C65} (2002) 017001.\vs
\bibitem{evgeni} E. Epelbaum, W. Gl\"ockle and Ulf-G. Mei{\ss}ner, {\it  Nucl. 
Phys.} {\bf A747} (2005) 362.\vs
\bibitem{fujita} J. Fujita and H. Miyawawa,  {\it Prog. Theor. Phys.} {\bf 17} 
(1957) 366.\vs
\bibitem{landau2nd} J.W. Holt, N. Kaiser and W. Weise, nucl-th/1106.5702.\vs
\end{thebibliography}
\end{document}